\documentclass[preprint,showpacs,preprintnumbers,amsmath,amssymb,nofootinbib]{revtex4}
\usepackage{booktabs}
\usepackage{mathrsfs}
\usepackage{epsfig}
\usepackage{graphicx}
\usepackage{dcolumn}
\usepackage{bm}
\usepackage{amsmath}
\usepackage{slashed}
\usepackage{multirow}
\usepackage{setspace}

\def\bit{\begin{itemize}}
\def\eit{\end{itemize}}
\def\ben{\begin{enumerate}}
\def\een{\end{enumerate}}
\def\bed{\begin{description}}
\def\eed{\end{description}}
\def\b{\beta}

\def\k{\kappa}
\def\l{\lambda}

\def\cmg{\, {\rm cm^2/g} }

\def\lsim{\raise0.3ex\hbox{$<$\kern-0.75em\raise-1.1ex\hbox{$\sim$}}}
\def\gsim{\raise0.3ex\hbox{$>$\kern-0.75em\raise-1.1ex\hbox{$\sim$}}}

\let\jnfont=\rm
\def\NPB#1,{{\jnfont Nucl.\ Phys.\ B }{\bf #1},}
\def\PLB#1,{{\jnfont Phys.\ Lett.\ B }{\bf #1},}
\def\EPJC#1,{{\jnfont Eur.\ Phys.\ Jour.\ C }{\bf #1},}
\def\PRD#1,{{\jnfont Phys.\ Rev.\ D }{\bf #1},}
\def\PRL#1,{{\jnfont Phys.\ Rev.\ Lett.\ }{\bf #1},}
\def\MPLA#1,{{\jnfont Mod.\ Phys.\ Lett.\ A }{\bf #1},}
\def\JPG#1,{{\jnfont J.\ Phys.\ G}{\bf #1},}
\def\CTP#1,{{\jnfont Commun.\ Theor.\ Phys.\ }{\bf #1},}
\def\JHEP#1,{{\jnfont JHEP \ }{\bf #1},}
\def\NPPS#1,{{\jnfont Nucl.\ Phys.\ Proc.\ Suppl.\ }{\bf #1},}

\def\beq{\begin{equation}}
\def\eeq{\end{equation}}
\def\bea{\begin{eqnarray}}
\def\eea{\end{eqnarray}}
\newcommand{\ba}{\begin{array}}
\newcommand{\ea}{\end{array}}
\def\nn{\nonumber}

\begin{document}

\title{Higgs exotic decays in general NMSSM with self-interacting dark matter}

\author{Wenyu Wang$^1$}
\author{Mengchao Zhang$^2$}
\author{Jun Zhao$^{1,2}$}

\affiliation{$^1$ Institute of Theoretical Physics, College of Applied Science,
Beijing University of Technology, Beijing 100124, China\\
$^2$ Key Laboratory of Theoretical Physics, Institute of Theoretical Physics,
     Chinese Academy of Sciences, Beijing 100190, China
\vspace{1cm} }

\begin{abstract}
Under current LHC and dark matter constraints,
the general NMSSM can have self-interacting dark matter to
explain the cosmological small structure.
In this scenario, the dark matter is the light singlino-like neutralino ($\chi$)
which self-interacts through exchanging the light singlet-like scalars ($h_1, a_1$).
These light scalars and neutralinos inevitably interact with the 125 GeV SM-like Higgs
boson ($h_{SM}$), which cause the Higgs exotic decays  $h_{SM} \to h_1 h_1, a_1 a_1, \chi\chi$.
We first demonstrate the parameter space required by the explanation of
 the cosmological small structure and then display the Higgs exotic decays.
We find that in such a parameter space the Higgs exotic decays can have
branching ratios of a few percent, which should be accessible in
the future $e^+e^-$ colliders.
\end{abstract}

\maketitle

\section{Introduction}
The $\Lambda$CDM cosmological model ($\Lambda$ Cold Dark Matter) can successfully describe
the large-scale structure of the Universe ($> 1$ Mpc) and the Cosmic Microwave Background
(CMB). Despite the success in describing the large-scale structure,
several recent observations at galactic or smaller scales fail to be explained
\cite{Bringmann:2013vra}, such as the problems of
\textit{missing satellites} \cite{Klypin:1999uc},
\textit{cusp vs ~core} \cite{deNaray:2011hy}
and \textit{too big to fail} ~\cite{BoylanKolchin:2011de}.
These issues strongly indicate that dark matter (DM) should not be composed of cold collisionless
particles, and instead may have a richer structure involving nontrivial self-interactions.
In particular, the self-interacting dark matter with a light force carrier ($\lsim$ 100 MeV)
may have a non-trivial velocity-dependent scattering cross section, which gives rise to the
appropriate cross sections for all of the small scales (the dwarf size, the Milky Way size
as well as the galaxy cluster size) \cite{Tulin:2012wi,Ko:2014nha,Laha:2013gva}. Note that
there could be small-scale power suppression due to self-interacting dark radiation
(which could also lead to dark acoustic oscillations) and/or late-kinetic decoupling.
\cite{darkradiation}

Such a self-interacting dark matter can be naturally achieved in the general singlet extension
of the MSSM (GNMSSM) \cite{fayet,NMSSM,Kaminska:2014wia}. In the GNMSSM, the SUSY preserving
$\mu$-term is dynamically generated by the vacuum expectation values (VEV) of a singlet
superfield $S$, which can solve the notorious $\mu$-problem in the MSSM. \cite{muew}
In the small $\lambda$ limit, the singlet sector can be almost decoupled from the eletroweak
symmetry breaking sector and becomes a dark sector in the theory. Due to its very weak
interactions with the SM particles, the $O (1)$ GeV singlino-like dark matter dominantly
annihilates into the light singlet-like scalars, which
can correctly produce the DM relic density and a proper Sommerfeld enhancement
factor \cite{NMSSM2,Hooper:2009gm,Wang:2009rj}. Such a feature can be used to solve the
small cosmological scale anomaly without conflicting with dark matter direct
detection \cite{Wang:2014kja}. Besides, the light singlet-like scalars are essential in
such a GNMSSM explanation of small structure problem. This will inevitably lead to various
exotic decays of the 125 GeV SM-like Higgs boson, such as the invisible decay to dark matter
and the decays to a pair of light scalars. \cite{Nhung:2013lpa}
Therefore, search for these exotic decays at colliders can allow for a test
of the self-interacting dark matter in the GNMSSM.

In this work, we first revisit the GNMSSM explanation of the small structure problem
under current LHC data and then investigate the 125 GeV Higgs exotic decays
in the allowed parameter space.
We organize the content as follows. In Sec. \ref{sec2}, we describe the general
self-interacting dark matter interactions for explaining the small cosmological
scale anomaly. In Sec. \ref{sec3} we briefly review the GNMSSM and in  Sec. \ref{sec4}
we demonstrate the parameter space of the GNMSSM
required by the explanation of the small cosmological scale anomaly.
In Sec. \ref{sec5},  we show the Higgs exotic decays in such a parameter space.
Sec. \ref{sec6} contains our conclusions.

\section{Self-interacting dark matter for small scale structure}\label{sec2}
\begin{figure}
\begin{center}
\scalebox{1.0}{\epsfig{file=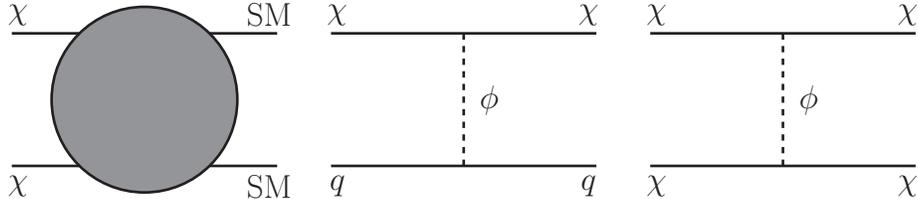}}\end{center}
\vspace{-0.7cm}
\caption{Dark matter interactions: annihilation to SM particles (left panel),
scattering off quarks (middle panel), and self-scattering process (right panel).}
\label{fig1}
\end{figure}

To explain both large scale and small scale structures of the Universe,
one can introduce the self-interacting DM scenario, in which the interactions
between DM and SM particles
can be summarized as (shown in Fig. \ref{fig1}):
\begin{enumerate}
\item The cross section of the DM annihilation to the SM particles
(the left diagram of Fig. \ref{fig1}), which at high energy
      determines the relic density of dark matter.
At low energy this cross section can be probed by the indirect detection
      experiments like PAMELA \cite{pamela} and AMS02 \cite{AMS-2013}.
When the early universe was cooling down, the equilibrium between DM and SM particles
in the thermal bath can no longer be maintained. The DM will annihilate to SM particles
until the annihilation rate falls below the expansion rate of the Universe.
In our following calculation, we use the standard method \cite{susy-dm-review} to calculate
the relativistic annihilation cross section and the degrees of freedom at the freeze-out
temperature.
\item The direct DM elastic scattering off the SM particles (the middle diagram of Fig. \ref{fig1}),
 which can be probed by various underground detection experiments, such as CDMS \cite{cdms2},
 XENON \cite{XENON100} and LUX \cite{lux}. Due to the high sensitivity of the
spin-independent (SI) measurement, we only consider the SI elastic interactions of DM
(denoted by $\chi$) and nucleon (proton and neutron), which are dominantly induced by
the scalar mediator at tree level, as shown in the middle diagram of Fig.(\ref{fig1}).
\item The non-relativistic self-scattering (the right diagram of Fig. \ref{fig1}),
where $\ell=0$ in the partial wave expansion gives the Sommerfeld enhancement relative
to the relativistic annihilation. Such an enhancement was first proposed to explain some
DM indirect detection results, like the positron excess observed by PAMELA
or AMS02.\cite{sommerfeld2} On the other hand, when $\ell \lsim 25$, the anomalies in the small
cosmological scales can be successfully accounted for.
\end{enumerate}
To explain the DM indirect detection results (such as the PAMELA result) and the small
cosmological scale anomalies, the DM scattering cross section should be calculated at
low energies, where the above interactions entangles with each other. In this case,
the complete investigation of the DM properties needs to consider all the above interactions
under various dark matter detection experiments. Details of this study can be found
in \cite{Wang:2014kja}. Here we only briefly discuss the calculation of the
non-relativistic self-scattering cross section.

The numerical input for the simulation of small scales is the differential cross section
$d \sigma/ d \Omega$, which is a function of the DM relative velocity $v$.
Then the viscosity (or conductivity) cross section $\sigma_V$ ~\cite{krstic:1999} can
be defined as
\beq
\sigma_V =  \int d \Omega \, \sin^2 \theta \, \frac{d\sigma}{d\Omega} \, .  \label{sigmaT}
\eeq
This formula is valid for the Majorana-fermion DM candidate. The weight of $\sin^2 \theta$ in
the integral is needed since both forward and backward scatterings amplitudes will diverge.
Such singularities correspond to the poles in
the $t$- and $u$-channel diagrams for the identical DM candidate.
Within the resonance region, $\sigma_V$ must be computed by solving the Schr\"{o}dinger
equation with the partial wave expansion method:
\beq
\frac{1}{r^2} \frac{d}{dr} \Big( r^2 \frac{d R_{\ell}}{dr} \Big)
+ \Big( k^2 - \frac{\ell (\ell + 1)}{r^2} - 2m_r V(r) \Big) R_\ell = 0. \label{radial}
\eeq
The total wave function of the spin-1/2 fermionic DM must be antisymmetric with respect to
the interchange of two identical particles.
Then the spatial wave function should be symmetric (antisymmetric) when the total spin is 0 (1).
The viscosity cross section can be reduced to two variables:
\bea
\frac{d\sigma_{VS}}{d\Omega} &=& \left|f(\theta)+f(\pi-\theta)\right|^2 = \frac{1}{k^2}\left|
\sum_{\ell (\mbox{\tiny\rm EVEN\  number})}^{\infty} (2\ell + 1) (\exp(2i\delta_l)-1)P_\ell(\cos\theta)\right|^2
 \label{sigmaVSsum}\\
\frac{d\sigma_{VA}}{d\Omega} &=& \left|f(\theta)-f(\pi-\theta)\right|^2 = \frac{1}{k^2}\left|
\sum_{\ell (\mbox{\tiny\rm ODD\ number})}^{\infty} (2\ell + 1) (\exp(2i\delta_l)-1)P_\ell(\cos\theta)\right|^2
\label{sigmaVAsum}
\eea
Using the orthogonality relation of the Legendre polynomials, we can have
\bea
\frac{\sigma_{VS}k^2}{4\pi} &=& \sum_{\ell(\mbox{\tiny EVEN\ number})}^{\infty}4\sin^{2}(\delta_{\ell+2}
-\delta_{\ell})(\ell+1)(\ell+2)/(2\ell+3) ,\label{sigmaVS}\\
\frac{\sigma_{VA}k^2}{4\pi} &=& \sum_{\ell(\mbox{\tiny ODD\ number})}^{\infty}4\sin^{2}(\delta_{\ell+2}
-\delta_{\ell})(\ell+1)(\ell+2)/(2\ell+3) . \label{sigmaVA1}
\eea
When the partial wave $\ell$ grows up, it can be seen that $\sigma_V$ will converge to
a static value as the phase shift $\delta_\ell$ approaches to a same value. We adopt the
numerical method in \cite{Tulin:2013teo} to calculate all these cross sections.
In our following analysis, we assume that the DM scatters randomly. Thus the average cross
section will be
\bea
\sigma_{V} =\frac{1}{4}\sigma_{VS}+\frac{3}{4} \sigma_{VA}\; .\label{sigmaVa;;}
\eea

\section{The general singlet extension of MSSM (GNMSSM)}\label{sec3}
The superpotential of the general NMSSM is given by \cite{Ellwanger:2009dp}
\beq\label{2.1e}
W=W_\mathrm{Yukawa} +(\mu + \lambda \widehat{S})\,\widehat{H}_u \cdot
\widehat{H}_d + \xi_F \widehat{S} + \frac{1}{2} \mu' \widehat{S}^2 +
\frac{\kappa}{3} \widehat{S}^3
\eeq
where $\widehat{H}_u$ and $\widehat{H}_d$ are the two Higgs doublets, $\widehat{S}$ is the
gauge singlet. The terms $\sim \mu$, $\mu'$ are supersymmetric mass terms, and $\xi_F$ of
dimension $mass^2$ parameterizes the (supersymmetric) tadpole term. The Yukawa couplings
of the quark and lepton superfields are given as
\beq\label{2.2e}
W_\mathrm{Yukawa} = h_u\, \widehat{Q} \cdot \widehat{H}_u\;
\widehat{U}^c_R + h_d\, \widehat{H}_d \cdot \widehat{Q}\;
\widehat{D}^c_R + h_e\, \widehat{H}_d \cdot \widehat{L}\;
\widehat{E}_R^c
\eeq
where the Yukawa couplings $h_u$, $h_d$, $h_e$ and the superfields
$\widehat{Q}$, $\widehat{U}^c_R$, $\widehat{D}^c_R$, $\widehat{L}$ and
$\widehat{E}_R^c$ should be understood as matrices and vectors in
family space, respectively. The corresponding soft SUSY breaking
masses and couplings are
\bea\label{2.5e}
-{\cal L}_\mathrm{soft} &=&
m_{H_u}^2 | H_u |^2 + m_{H_d}^2 | H_d |^2
+ m_{S}^2 | S |^2+m_Q^2|Q^2| + m_U^2|U_R^2| \nn \\
&&+m_D^2|D_R^2| +m_L^2|L^2| +m_E^2|E_R^2|
\nn \\
&&+ (h_u A_u\; Q \cdot H_u\; U_R^c - h_d A_d\; Q \cdot H_d\; D_R^c
- h_{e} A_{e}\; L \cdot H_d\; E_R^c\nn \\ &&
+\lambda A_\lambda\, H_u \cdot H_d\; S + \frac{1}{3} \kappa A_\kappa\,
S^3 + m_3^2\, H_u \cdot H_d + \frac{1}{2}m_{S}'^2\, S^2 + \xi_S\, S
+ \mathrm{h.c.}) \; .
\eea
Clearly, if $\mu = \mu' = \xi_F = 0$, the $\mathcal{Z}_3$ NMSSM
superpotential is obtained as
\beq\label{2.6e}
W_\mathrm{NMSSM} = \lambda \widehat{S}\,\widehat{H}_u \cdot
\widehat{H}_d + \frac{\kappa}{3} \widehat{S}^3
\eeq
and the parameters $m_3^2$, $m_{S}'^2$ and $\xi_S$ in
(\ref{2.5e}) are vanishing. Then, a VEV $s$ of $\widehat{S}$ of the order of
the weak or SUSY breaking scale generates an effective $\mu$-term with
\beq\label{2.7e}
\mu_\mathrm{eff} = \l s\;
\eeq

The neutral physical Higgs fields (with index $R$ for the CP-even and index $I$
for the CP-odd states) can be obtained by expanding the full scalar potential
 around the real neutral vevs $v_u$, $v_d$ and $s$
\beq\label{2.10e}
H_u^0 = v_u + \frac{H_{uR} + iH_{uI}}{\sqrt{2}} , \quad
H_d^0 = v_d + \frac{H_{dR} + iH_{dI}}{\sqrt{2}} , \quad
S = s + \frac{S_R + iS_I}{\sqrt{2}}\; .
\eeq
It is convenient to define, together with $\mu_\mathrm{eff}$
as in (\ref{2.7e}),
\beq\label{2.14e}
B_\mathrm{eff} = A_\lambda+ \kappa s,
\quad \widehat{m}_3^2 = m_3^2 + \lambda(\mu' s + \xi_F)
\eeq
where we have used the convention $\mu=0$. $B_\mathrm{eff}$ plays the role of
the MSSM-like $B$-parameter, and $\widehat{m}_3^2$ vanishes
in the NMSSM. After the spontaneously symmetry breaking,
we can have the mass spectrum of all the particles.
The tree level Higgs mass matrices can be obtained by expanding the full
Higgs potential around the real neutral vevs $v_u$, $v_d$ and $s$ as in (\ref{2.10e}).
Then, the elements of the $3 \times 3$ CP-even mass matrix ${\cal
M}_S^2$ in the basis $(H_{dR}, H_{uR}, S_R)$ after the elimination
of $m_{H_d}^2$, $m_{H_u}^2$ and $m_{S}^2$ can be read as
\bea
{\cal M}_{S,11}^2 & = & g^2 v_d^2 + (\mu_\mathrm{eff}\, B_\mathrm{eff} +
\widehat{m}_3^2)\,\tan\beta\;, \nn\\
{\cal M}_{S,22}^2 & = & g^2 v_u^2 + (\mu_\mathrm{eff}\, B_\mathrm{eff} +
\widehat{m}_3^2)/\tan\beta\;, \nn\\
{\cal M}_{S,33}^2 & = & \l (A_\l + \mu') \frac{v_u v_d}{s}
+ \k s (A_\k + 4\k s+ 3 \mu') - (\xi_S + \xi_F \mu')/s\;, \nn\\
{\cal M}_{S,12}^2 & = & (2\l^2 - g^2) v_u v_d -
\mu_\mathrm{eff}\, B_\mathrm{eff} - \widehat{m}_3^2 \;, \nn\\
{\cal M}_{S,13}^2 & = & \l (2 \mu_\mathrm{eff}\, v_d -
(B_\mathrm{eff} + \k s + \mu')v_u)\;, \nn\\
{\cal M}_{S,23}^2 & = & \l (2 \mu_\mathrm{eff}\, v_u -
(B_\mathrm{eff} + \k s + \mu')v_d)\;.
\label{2.22e}
\eea
After diagonalization, we can have three CP-even mass eigenstates, which are denoted
as $h_1$, $h_2$, $h_3$, respectively. The element of the CP-odd mass matrix
${\cal M'}_P^2$  is $3 \times 3$ matrix in the basis $(H_{dI}, H_{uI}, S_I)$. ${\cal M'}_{P}^2$
usually contains a massless Goldstone mode ${G}$. We rotate this mass matrix into the basis
(${A}, {G},S_I$), where ${A} = \cos\b\, H_{uI}+ \sin\b\, H_{dI}$.
After absorbing the Goldstone mode, the remaining $2 \times 2$ mass matrix ${\cal M}_{P}^2$
in the basis ($A, S_I$) has the following elements
\bea
{\cal M}_{P,11}^2 & = & \frac{2 (\mu_\mathrm{eff}\, B_\mathrm{eff} +
\widehat{m}_3^2)}{\sin 2\b}\; , \nn\\
{\cal M}_{P,22}^2 & = & \l (B_\mathrm{eff}+3\k s +\mu')\frac{v_u
v_d}{s} -3\k A_\k s  -2 m_{S}'^2 -\k \mu' s
-\xi_F\left(4\k + \frac{\mu'}{s}\right) -\frac{\xi_S}{s}\; , \nn\\
{\cal M}_{P,12}^2 & = &\l (A_\l - 2\k s - \mu')\, v\;.
\label{2.26e}
\eea
Similarly, we can obtain two mass eigenstates denoted as ($a_1$, $a_2$). Note that we can set
$A_\l - 2\k s - \mu'=0$ to forbid the mixing between the doublet and the singlet.
In the neutralino sector, the neutral gaugino $\l_1$ and $\l_2^3$ mix with the neutral
higgsinos $\psi_d^0, \psi_u^0, \psi_S$, which generates a symmetric $5 \times 5$ mass
matrix ${\cal M}_0$. In the basis $\psi^0 = (-i\l_1 , -i\l_2^3, \psi_d^0, \psi_u^0, \psi_S)$,
the resulting mass terms in the Lagrangian can be read as
\beq\label{2.31e}
{\cal L} = - \frac{1}{2} (\psi^0)^T {\cal M}_0 (\psi^0) + \mathrm{h.c.}
\eeq
where
\beq\label{2.32e}
{\cal M}_0 =
\left( \ba{ccccc}
M_1 & 0 & -\frac{g_1 v_d}{\sqrt{2}} & \frac{g_1 v_u}{\sqrt{2}} & 0 \\
& M_2 & \frac{g_2 v_d}{\sqrt{2}} & -\frac{g_2 v_u}{\sqrt{2}} & 0 \\
& & 0 & -\mu_\mathrm{eff} & -\l v_u \\
& & & 0 & -\l v_d \\
& & & & 2 \k s + \mu'
\ea \right),
\eeq
Here $M_1,~M_2$ are the soft mass parameters of the gauginos. After diagonalization, we can have
five mass eigenstates $\chi_{1,\cdots,5}$. If the neutralino ($\chi_1$) is the lightest
supersymmetric pariticle (LSP), it will be a good candidate of DM. In the GNMSSM, the
self-interaction dark matter scenario can be realized if the LSP is singlino-like and $h_1$ is
light enough and serve as be the mediator between the DM scattering \cite{Wang:2014kja}.
In the following section we will revisit such a scenario under the current LHC constraints
and investigate the the Higgs exotic decays in the visible parameter space of explaining
the small structure anomalies.

\section{Self-interacting dark matter in the GNMSSM}\label{sec4}
From the neutralino mass matrix in Eq.(\ref{2.32e}), we can see that the GNMSSM singlino
mass is mainly determined by the parameters $\mu'$ and $\kappa$. By tuning these two parameters,
the very light singlino-like LSPs can dominantly annihilate to the singlet-like Higgs bosons
and satisfy the requirements of relic abundance and direct detections \cite{Wang:2014kja}.
Here it should be mentioned that the correlation between the singlino sector and the singlet
Higgs bosons sector are relaxed by the extra parameter $\mu'$ in the GNMSSM, which is different
from the situation in ${\cal Z}_3$-NMSSM \footnote{In the ${\cal Z}_3$-NMSSM, the masses of the
singlino and the singlet Higgs bosons are strongly related by the same parameter $\kappa$.
Thus, when $\kappa \ll \lambda$, \cite{Draper:2010ew}
the correct relic abundance of the very light singlino
dark matter is obtained by annihilating to the SM fermions via the very light (several MeV)
singlet CP-odd Higgs boson. However, such a light singlet scalar will greatly enhance the
DM-nucleon SI cross section and is disfavored by the current dark matter direct detection.
So, it hardly realizes the self-scattering dark matter for the cosmological small structure in
the ${\cal Z}_3$ NMSSM.}

In our numerical calculations, we use the newest version of package NMSSMTools \cite{nmssmtools}
to perform the scan of the parameter space. All the mass input parameters are defined at the
electroweak scale. We choose the input parameters of the Higgs sector as
\bea\label{2.17e}
\lambda,\ \kappa,\ \tan\beta,\ \mu_\mathrm{eff},\ A_\lambda,\ A_\kappa.\
 m_3^2,\ \mu',\ m_{S}'^2,\ \xi_F,\ \xi_S.
\eea
Then, we scan the dimensionless parameters in the range
$$0.005<|\lambda,\ \kappa|<0.5,~~1<\tan\beta<50$$
and all the other mass dimension parameters in the range
$$(-2{\rm TeV},~2{\rm TeV}).$$
To obtain the very light singlino-like DM and the singlet-like Higgs bosons, we require
\begin{eqnarray}
\xi_S&\sim & \l (A_\l + \mu') v_u v_d - \k s^2 (A_\k + 4\k s+ 3 \mu') -  \xi_F \mu', \label{eq:req1}\\
m_{S}'^2& \sim & \frac{1}{2}\left[\l (B_\mathrm{eff}+3\k s +\mu')\frac{v_u
v_d}{s} -3\k A_\k s    -\k \mu' s
-\xi_F\left(4\k + \frac{\mu'}{s}\right) -\frac{\xi_S}{s}\right],  \label{eq:req2}  \\
 \mu' &\sim& -2 \k s .  \label{eq:req3}
\end{eqnarray}
Also, we require
\begin{eqnarray}
  \label{eq:amx}
  A_\l= 2\k s +\mu',
\end{eqnarray}
to forbid the mixing between the CP-odd doublet and the singlet Higgs bosons.
The gaugino input parameters $M_1,~M_2,~M_3$ are scanned in the range of
$(-1 {\rm TeV},~1 {\rm TeV})$. In order to give a 125 GeV SM Higgs boson, we scan the third
generation squark mass parameters $m_{Q_{3L}}$, $m_{U_{3R}}$ and $m_{D_{3R}}$ in the range of
$(-5{\rm TeV},~5{\rm TeV})$. The first two generation squark and slepton mass parameters
are fixed at 5 TeV. Finally, we tune the parameter $A_\k$ slightly to give a light $h_1$ at
order of 10 MeV. In our scan, we impose the following constraints:
\begin{itemize}
\item The SM-like Higgs mass in the range of 123-127 GeV;
\item The thermal relic density of the lightest neutralino in the 2$\sigma$ range of the
Planck value \cite{planck};
\item The LEP-I bound on the invisible $Z$-decay, $\Gamma(Z \to \tilde{\chi}^0_1 \tilde{\chi}^0_1) < 1.76$ MeV, and the LEP-II upper bound on $\sigma(e^+e^- \to\tilde{\chi}^0_1 \tilde{\chi}^0_i) <5 \times 10^{-2}~{\rm pb}$ for $i>1$, as well as the lower mass bounds on the sparticles from the direct searches at LEP and the Tevatron;
\item The constraints from the LHEP-II direct search for the Higgs boson exotic decays, including the decay modes $h \to h_1 h_1, a_1 a_1 \to 4 f$;
\item The constraints from $B$-physics: $B \to X_s \gamma$, $B^+ \to \tau^+ \nu$, $\Upsilon \to \gamma a_1 $, the $a_1$--$\eta_b$ mixing and the mass difference $\Delta M_d$ and $\Delta M_s$.
\end{itemize}

\begin{figure}
\begin{center}\scalebox{1}{\epsfig{file=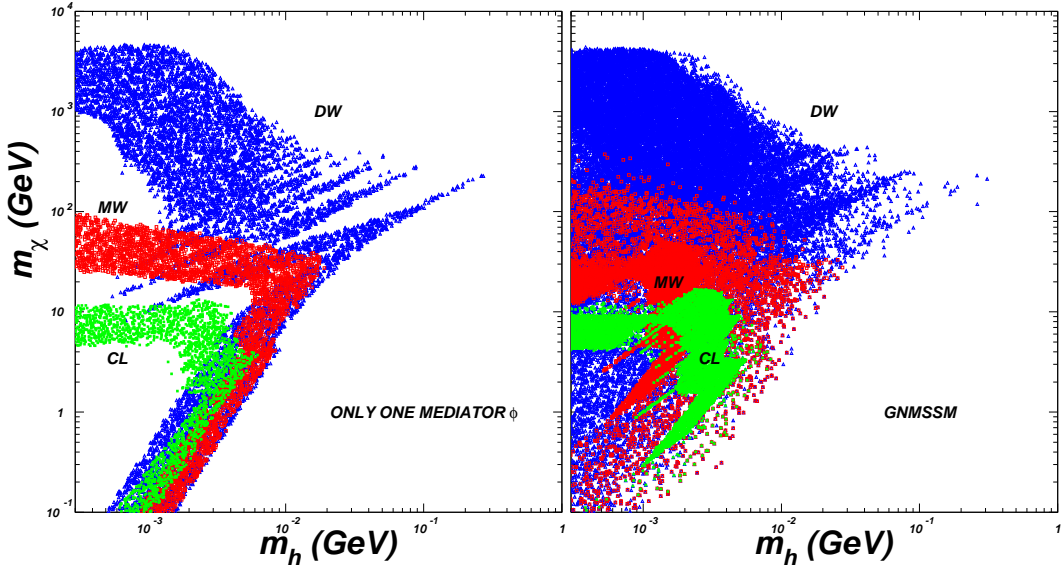}}\end{center}
\caption{The survived points under the constraints of relic density and the scattering
cross section in case of $\l=0$. The blue points are the points satisfy the simulation
in the Dwarf scale
($\sigma/m_\chi \sim 0.1 - 10 \, \cmg$, the characteristic velocity is 10 km/s.)
The red points are the points satisfy the simulation in the Milky Way
($\sigma/m_\chi \sim 0.1 - 1 \, \cmg$, the characteristic velocity is 200 km/s.)
The green points are the points satisfy the simulation in the Milky Way
($\sigma/m_\chi \sim 0.1 - 1 \, \cmg$, the characteristic velocity is 1000 km/s.)}
\label{fig2}
\end{figure}
In Fig.~\ref{fig2}, we show the survived points under the constraints of relic density and
scattering cross section in case of $\l=0$. The left panel shows the results for one
mediator in the simplified model \cite{Kang:2016xrm}, while the right panel shows
the results of the GNMSSM. We can see that the simulation of small structure gives a
stringent constraint on the parameter space for the self-interacting DM, in particular
for one light mediator case. The survival parameter space in the GNMSSM is larger than
the simple model of one mediator. The main reason is that in the DM self-interaction
model \cite{Tulin:2013teo} DM can only annihilate into $hh$ via $t$-channel and $u$-channel
while in the GNMSSM DM can annihilate into $hh$, $ha$ and $aa$ via $t$-channel, $u$-channel
and $s$-channel.

\begin{figure}
\begin{center}
\epsfig{file=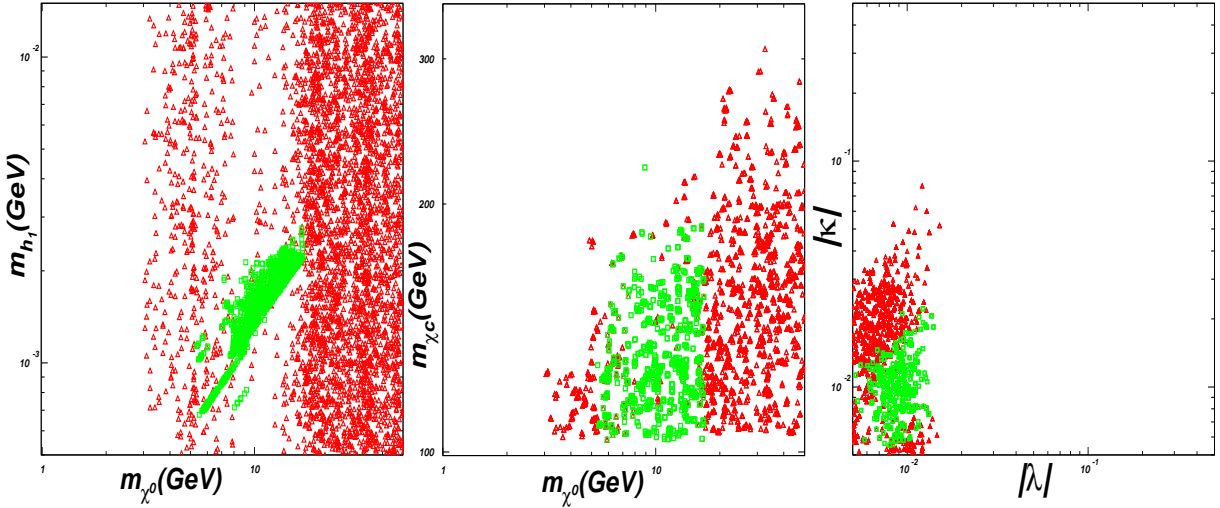,width=16cm,height=7cm}
\end{center}
\caption{Scatter plots of the samples which satisfy the current experimental
constraints except the direct detection limits of dark matter.
The $\Box$ (green) and  $\triangle$ (red) samples can and cannot
explain the small-scale cosmic structure, respectively.}
\label{fig3}
\end{figure}
\begin{figure}[h]
\begin{center}
\scalebox{0.5}{\epsfig{file=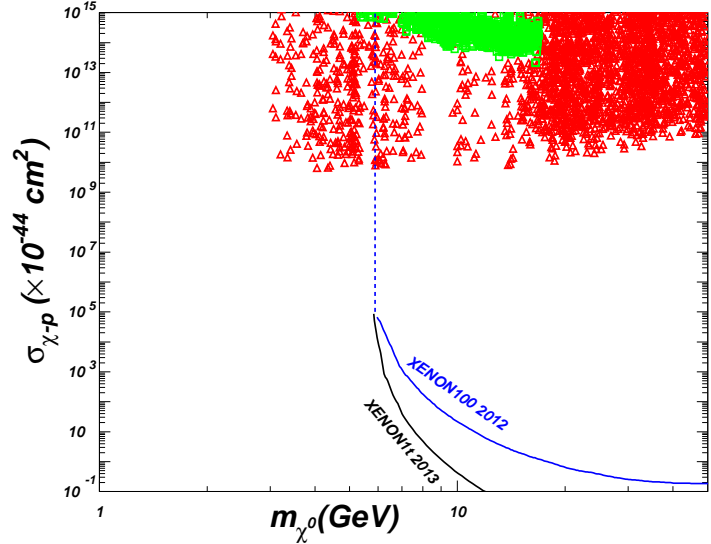}}
\end{center}
\caption{Same as Fig. \ref{fig3}, but showing the spin-independent cross section of
dark matter-nucleon scattering.
The curves are the current limits of dark matter direct detections,
which can exclude most samples but have no sensitivity to ultra-light dark matter.}
\label{fig4}
\end{figure}
In Fig.~\ref{fig3}, we present the scatter plots
of the survived samples which can satisfy the current experimental
constraints and can explain the small-scale cosmic structure.
From Fig.~\ref{fig3}, it can be seen
that the mass of the light CP-even singlet-like Higgs boson has to be in the range
of (0.6, 3) MeV
in order to explain the small structure simulations. The mass of the singlino-like DM is
required to be in the range of (5, 11) GeV. Besides, the couplings $\l$ and $\k$ are
bounded below 0.02
and $\k$ can be much larger than $\l$. We also note that the chargino mass $m_{\chi^\pm_1}$ is
strongly constrained and should be less than 210 GeV.

In Fig. \ref{fig4}, we display the scatter plots by showing the spin-independent
cross section of dark matter-nucleon scattering.
We can see that the cross section is greatly enhanced by the light mediator $h_1$ with mass
about 10 MeV. The current direct detection experiments, such as XENON 2012,
have  already excluded most of the samples when the dark matter is not so light.
However, the ultra-light dark matter can still be allowed
because the direct detections have no sensitivity to
ultra-light dark matter.

\section{Higgs exotic decays in the GNMSSM with self-scattering dark matter}\label{sec5}
\begin{figure}[h]
\begin{center}
\epsfig{file=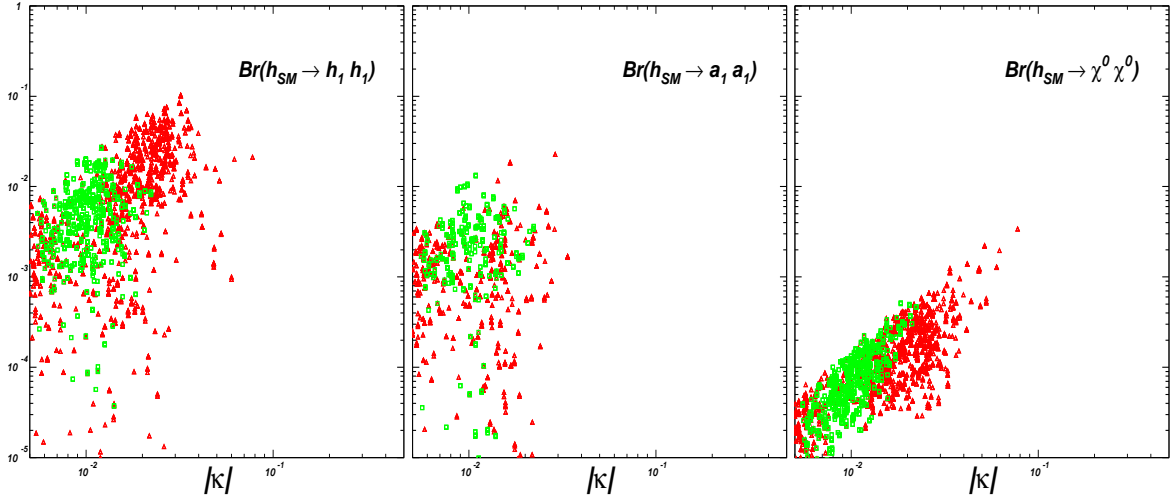,width=16cm,height=7cm}
\end{center}
\caption{Same as Fig. \ref{fig3},
but showing the branching ratios of the exotic decays
of the 125 GeV SM-like Higgs boson. }
\label{fig5}
\end{figure}

Now we display in Fig.~\ref{fig5} the Higgs exotic decays in the GNMSSM parameter space
which can satisfy the current experimental
constraints (including the direct detection limits of dark matter)
and can explain the small-scale cosmic structure.
From Fig. \ref{fig5} we can see that the branching ratios of these channels become larger
with the increase of $\k$. This is because that the branching ratios of these decay channels
are dominantly determined by the Higgs self-coupling parameter $\k$ for the fixed masses
of $h_1$, $a_1$ and $\chi_1$. However, it should be noted that the small structure
simulations require a small $\k$ so that the branching ratios of the decay channels
$h_{SM} \to h_1 h_1, a_1 a_1$ can only reach about 3\%, while the invisible decay can
only reach 0.5\%.

\begin{figure}[h]
\begin{center}
\scalebox{0.5}{\epsfig{file=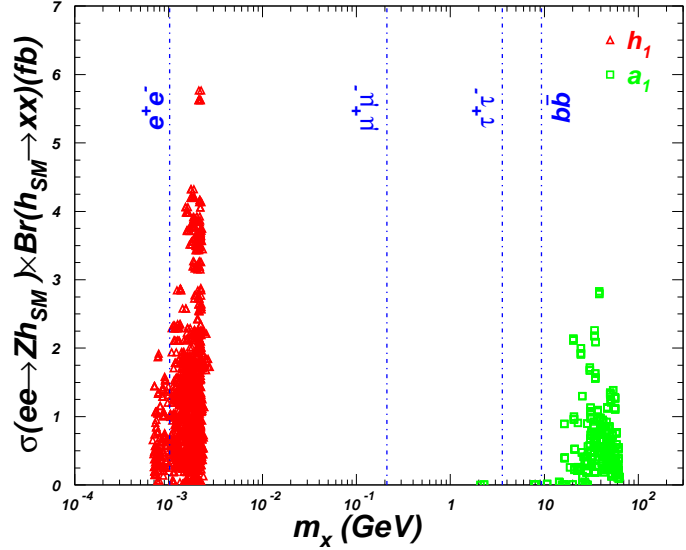}}
\end{center}
\caption{Same as Fig. \ref{fig3}, but showing the cross sections of
$\sigma(e^+e^- \to Zh_{SM})\times Br(h_{SM} \to XX)$ for $\sqrt{s}=250$ GeV,
where $X=h_1,a_1,\chi$.
The vertical lines (from left to right) correspond to the mass thresholds
of $e^+e^-$, $\mu^+ \mu^-$, $\tau^+\tau^-$ and $b\bar{b}$ final states, respectively.}
\label{fig6}
\end{figure}
In Fig.~\ref{fig6}, we show the cross sections of
$\sigma(e^+e^- \to Zh_{SM})\times Br(h_{SM} \to XX)$ ( $X=h_1,a_1,\chi$)
for $\sqrt{s}=250$ GeV.
From Fig.~\ref{fig6}, we can see that the cross section of $e^+e^- \to Zh_{SM} \to Z h_1 h_1$
can maximally reach  6 fb, which corresponds to $3 \times 10^4$ events for the highest
luminosity ${\cal L}=5000$ fb$^{-1}$ at an $e^+ e^-$ collider.
When $2m_e < m_{h_1}$, the CP-even $h_1$ will dominantly decay to the $e^+e^-$ final states.
Then the most promising channel of searching for such a light CP-even Higgs boson will be
$e^+e^- \to Zh_{SM} \to Z h_1 h_1 \to 2j+4\ell$ at an $e^+ e^-$ collider. While the cross section
of $e^+e^- \to Zh_{SM} \to Z a_1 a_1$ can maximally reach 3 fb and the light CP-odd $a_1$
will mainly decay to $b\bar{b}$, which leads to a distinctive signature of
$2\ell+4b$ at an $e^+ e^-$ collider.
\footnote{ The observability of light $a_1$ through the
process $pp \to Vh_{SM} \to 2\ell+4b$ has been investigated in \cite{Cheung:2007sva,han,zhang} and
is found to reach $3\sigma$ if $Br(h_{SM} \to a_1 a_1) > 15\%$ at LHC with a
luminosity ${\cal L}=300$ fb$^{-1}$ \cite{zhang}.}.
Due to the clean environment, with the mass window cut $|m_{4b}-m_{h_{SM}}|<10$ GeV,
such a signature may be observed at a future $e^+e^-$ collider.

\section{Conclusions}\label{sec6}
We revisited the general NMSSM which
can have self-interacting dark matter to
explain the cosmological small structure.
In this scenario, the dark matter is the ultra-light singlino-like neutralino ($\chi$)
which self-interacts through exchanging the ultra-light singlet-like scalars ($h_1, a_1$).
Since these light scalars and neutralinos inevitably interact with the 125 GeV SM-like Higgs
boson ($h_{SM}$), the Higgs will have exotic decays  $h_{SM} \to h_1 h_1, a_1 a_1, \chi\chi$.
We first showed the parameter space required by the explanation of
 the cosmological small structure and then displayed the Higgs exotic decays.
We found that in such a parameter space the Higgs exotic decays can have
branching ratios of a few percent, which could be accessible in
the future $e^+e^-$ colliders.

\section*{Acknowledgments}
We thank Lei Wu and Jin Min Yang for discussions and comments.
This work was supported by the Natural Science Foundation of China under grant number 11375001,
and by the talent foundation of education department of Beijing.

\end{document}